\documentclass[preprint]{vgtc}               

\ifpdf
  \pdfoutput=1\relax                   
  \usepackage{graphicx}                
  \DeclareGraphicsExtensions{.pdf,.png,.jpg,.jpeg} 
\else
  \usepackage{graphicx}                
  \DeclareGraphicsExtensions{.eps}     
\fi%

\graphicspath{{figures/}{pictures/}{images/}{./}} 

\usepackage{microtype}                 
\PassOptionsToPackage{warn}{textcomp}  
\usepackage{textcomp}                  
\usepackage{mathptmx}                  
\usepackage{times}                     
\usepackage{cite}                      
\usepackage{tabu}                      
\usepackage{booktabs}                  
\usepackage{placeins}
\usepackage{xcolor}

\onlineid{1042}

\vgtccategory{Application}

\vgtcinsertpkg

\title{Topological Analysis of Magnetic Reconnection in Kinetic Plasma Simulations}

\author{Divya Banesh \thanks{e-mail: dbanesh@ucdavis.edu}\\ %
         \parbox{1.4in}{\scriptsize \centering University of California Davis \\ Los Alamos National Lab}
\and Li-Ta Lo \\ %
     \scriptsize Los Alamos National Lab %
\and Patrick Kilian \\ %
     \scriptsize Los Alamos National Lab %
\and Fan Guo \\ %
     \scriptsize Los Alamos National Lab %
\and Bernd Hamann \\ %
     \scriptsize University of California Davis %
     }

\teaser{
 \includegraphics[trim=0 0 0 0,clip, width=.49\linewidth]{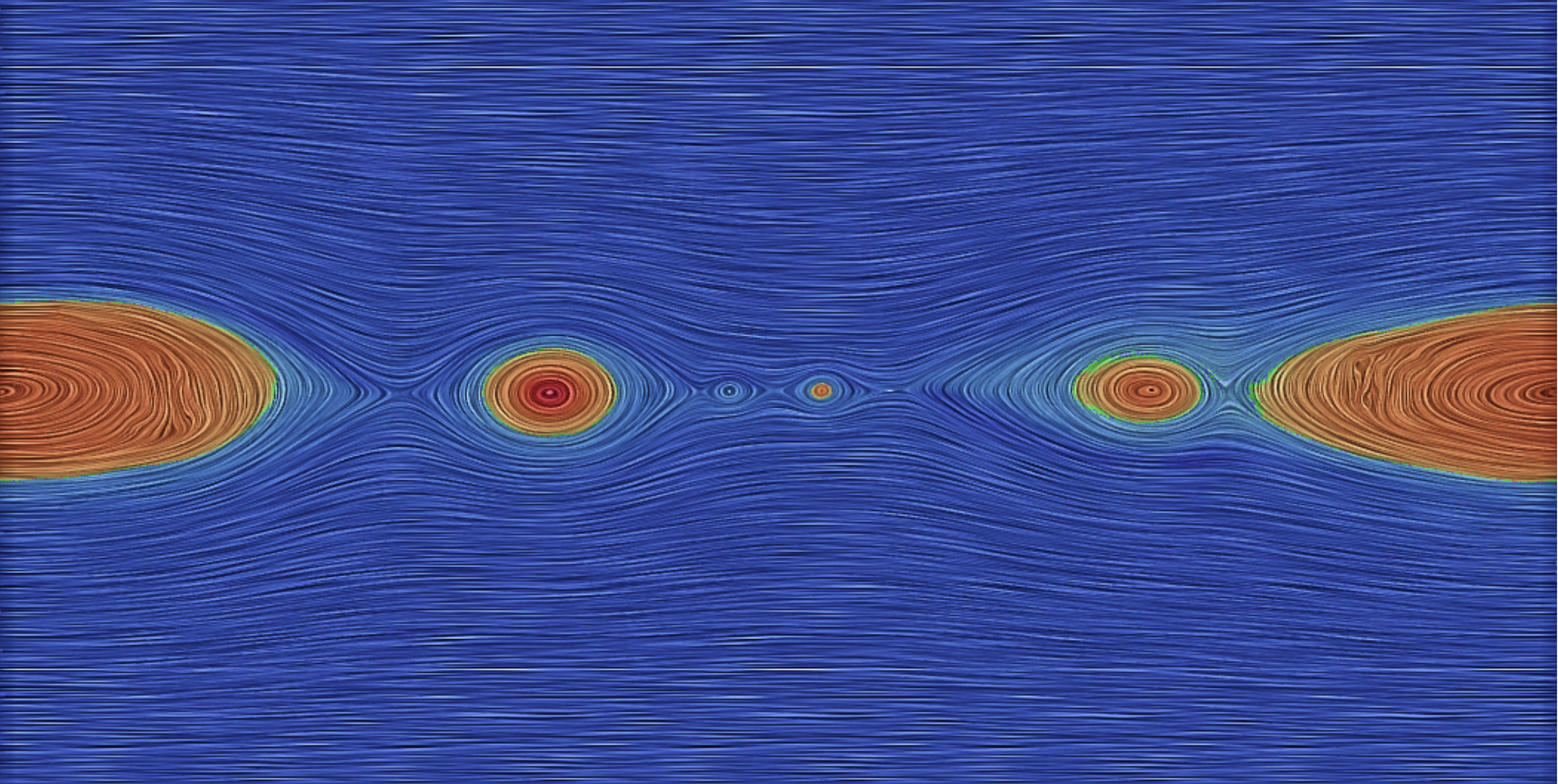}
 \includegraphics[trim=0 0 0 0,clip, width=.49\linewidth]{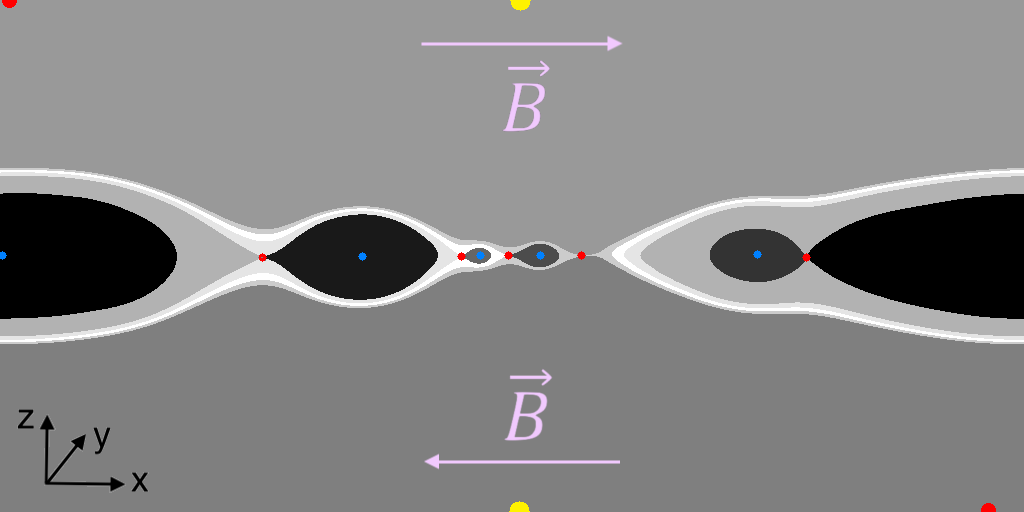}
 \vspace{-1em}
 \centering
  \caption{\textbf{Islands in Magnetic Reconnection.} A sample output of a two-dimensional Particle-in-Cell simulation. Left panel visualizes the in-plane magnetic field $B_\mathrm{x}$ and $B_\mathrm{z}$ using line integral convolution (LIC), colored by the out-of-plane component $B_\mathrm{y}$ with a standard hot-cold colormap. Right panel shows the results of contour-tree based segmentation on the y-component of magnetic vector potential, $A_\mathrm{y}$, derived from $\vec{B}$. Each segment is in a unique grey-toned color. In plasma physics, these segments are magnetic islands created by magnetic reconnection. The blue and yellow dots are, respectively, the maxima and minima of $A_\mathrm{y}$ while the red dots are saddle points. Physicists refer to minima and maxima as `O' points and saddles as `X' points. The X points correspond to the locations where magnetic field topology changes, i.e. reconnection occurs. A comparison of these figures shows clearly that regions of concentric circles in the LIC visualization are analogous to the segments defined by the contour-tree algorithm. In contrast to LIC, the contour-tree analysis precisely locates the reconnection (X) points and shows the nested hierarchy of the segments.}
\label{fig:teaser}
}

\abstract{Magnetic reconnection is a ubiquitous plasma process in which oppositely directed magnetic field lines break and rejoin, resulting in a change of the magnetic field topology. Reconnection generates magnetic islands: regions enclosed by magnetic field lines and separated by reconnection points. Proper identification of these features is important to understand particle acceleration and overall behavior of plasma. We present a contour-tree based visualization for robust and objective identification of islands and reconnection points in two-dimensional~(2D) magnetic reconnection simulations. The application of this visualization to a simple simulation has revealed a physical phenomenon previously not reported, resulting in a more comprehensive understanding of magnetic reconnection.%
} 

\CCScatlist{
  \CCScatTwelve{Human-centered computing}{Visu\-al\-iza\-tion}{Visu\-al\-iza\-tion techniques}{Treemaps};
  \CCScatTwelve{Human-centered computing}{Visu\-al\-iza\-tion}{Visualization design and evaluation methods}{}
}

\begin{document}


\firstsection{Introduction}

\maketitle

\begin{figure}[t!]
  \centering
  \includegraphics[width=\columnwidth]{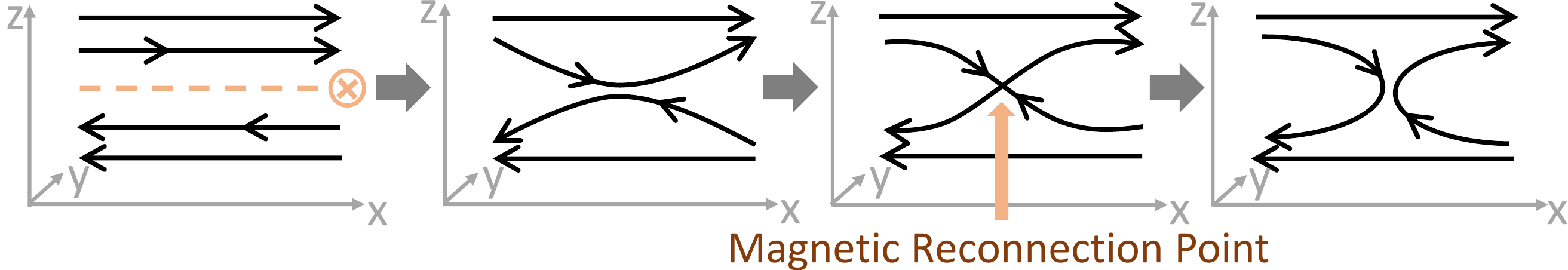}
  \vspace{-2em}
  \caption{\label{fig:magRecDiagram}
     \textbf{Magnetic reconnection process.} Oppositely directed field lines touch and break at the reconnection point. The dashed orange line shows the current sheet with electric current in the $y$ direction.}
     \vspace{-1.5em}
\end{figure}

In plasma, magnetic reconnection events change the magnetic field topology~\cite{Birn2001Geospace, Parker1957JGR, Biskamp1986Magnetic}. During the process, oppositely directed magnetic field lines bend towards each other and touch at a reconnection point. These field lines then break, pair and rejoin, as shown in Figure~\ref{fig:magRecDiagram}. This generates closed regions called magnetic islands and releases energy from the magnetic field into the plasma. Understanding reconnection holds the key for understanding high-energy particles in different plasma environments, such as Earth's magnetosphere \cite{Fu2011}, solar flares \cite{Chen2018}, and high-energy astrophysics \cite{Drenkhahn2002}.

Analysis of magnetic reconnection in three-dimensional~(3D) space requires the use of vector field topology. However, when restricting the analysis to a plane and thereby ignoring all variations perpendicular to it, we have determined that it is possible to use scalar field topology to identify features of interest. Since the magnetic field $\vec{B}$ is divergence-free ($\nabla \cdot \vec{B}$ = 0), it can be written as the curl of a magnetic vector potential $\vec{A}$
where $\nabla \times \vec{A} = \vec{B}$~\cite{griffiths1962introduction}.
%
%
%
%
We assume that $\vec{B}$ remains unchanged along the $y$ axis.
Therefore, for a simulation in the $x-z$~plane, the topologically relevant quantity is: 
 \vspace{-0.5em}
\[\nabla \times \vec{A_\mathrm{y}} = \left(\partial_\mathrm{x},0,\partial_\mathrm{z}\right)^T \times \left(0, A_\mathrm{y},0\right)^T = -\frac{\partial A_\mathrm{y}}{\partial z}\hat{\imath} + \frac{\partial A_\mathrm{y}}{\partial x}\hat{k}\quad.\]
\vspace{-0.5em}
Additionally, the gradient of $A_\mathrm{y}$ in the $x-z$~plane is
\[\nabla A_\mathrm{y} = \frac{\partial A_\mathrm{y}}{\partial x}\hat{\imath} + \frac{\partial A_\mathrm{y}}{\partial z}\hat{k}\quad.\] 
It is obvious that $(\nabla\times\vec{A_\mathrm{y}}) \cdot \nabla A_\mathrm{y} = 0$ in the $x-z$~plane, which means the gradient of $A_\mathrm{y}$ is perpendicular to the curl of $\vec{A_\mathrm{y}}$ and, equivalently, the in-plane magnetic field $\vec{B}$. The gradient of $A_\mathrm{y}$ is also perpendicular to the contours of $A_\mathrm{y}$ at regular, i.e., not critical, points. Therefore, the $\vec{B}$ field and the contours of $A_\mathrm{y}$ are parallel in the plane, making them topologically equivalent. Given this, we can apply topological analysis to the scalar field $A_\mathrm{y}$ in lieu of the vector field $\vec{B}$.


Particle-in-Cell~(PiC)~\cite{hockney1988computer} is a commonly used simulation method for magnetic reconnection. 
PiC simulations combine a mesh structure with numerous particles seeded in each cell. The mesh represents the electric and magnetic fields in the plasma, while the computational macro particles represent the charged particles. Similar to Monte Carlo methods, the number of particles used affects the noise level in the final result. This noise can be significantly reduced by increasing the number of particles in each cell, but is never completely eliminated. 

Despite the complexity of these simulations, until recently methods for analysis of magnetic reconnection simulations have been relatively simple. The traditional technique for reconnection visualization is a laborious process. It requires physicists to subjectively and iteratively parse through isovalues of the magnetic vector potential to find segments that are visually interesting. However, it is often true that the selected isovalues do not capture the segmentation entirely. Once segments are determined, corresponding critical points are approximated. There is no guarantee that this process identifies all islands or reconnection points.  

More recently, Servidio et al.~\cite{servidio2010statistics, servidio2009magnetic} presented a more rigorous algorithm for the visualization of magnetic islands and reconnection points. This method, developed for the analysis of fluid based magnetohydrodynamic~(MHD) simulations, calculates the gradient and Hessian matrix of the vector potential at each vertex. When the magnitude of the gradient is smaller than a user-defined threshold, the gradient is considered to be zero and the vertex is viewed as a critical point. Though this approach has produced sound results for MHD simulations, there are several problems that arise when applying it to PiC simulations. PiC simulations have a higher noise level compared to MHD simulations due to their stochastic nature. This noise produces inaccuracies when using the gradient-based algorithm. Haggerty et al.~\cite{haggerty2017exploring} discussed various techniques as a work-around to reduce the impact of noise to the results. Additionally, the user-defined threshold is an empirically-defined and subjective value with no physical correspondence to properties of the simulation. The implementation, developed at the University of Delaware, is proprietary and not easily obtainable. This has limited the usability of the algorithm, even within the physics community.

We use a contour-tree based segmentation algorithm for the objective identification of magnetic islands and reconnection points. Contour trees~\cite{carr2003computing} are widely used for data segmentation~\cite{rosen2017using, klacansky2019toward}. We use the open-source Topology Toolkit~(TTK)~\cite{tierny2017topology, gueunet2019task} for contour tree generation; it can easily be integrated into physicists' workflows. Our approach differs from other approaches due to the one-to-one correspondence between the mathematical definition of magnetic reconnection and 2D scalar field topology. The advantages of this technique for PiC simulations are demonstrated by applying the technique to an ensemble of 1000 evenly spaced, time-dependent data sets, produced as outputs from a single current sheet simulation. A major advantage of the contour-tree based algorithm is the use of persistence~\cite{edelsbrunner2010computational} to reduce the influence of noise on the segmentation. PiC noise is concentrated at short wavelengths in the $\vec{B}$ field. In the process of converting the $\vec{B}$ to $A_\mathrm{y}$ using the spectrum solver, this high-frequency component is smoothed. The remaining noise in $A_\mathrm{y}$ has a much lower amplitude compared to the dynamic range of $A_\mathrm{y}$; small regions with tall peaks (high persistence) cannot exist as a contribution of noise. Those remaining \textit{``small bumps''} can be removed by using a proper persistence value.
The persistence of a segment correlates with the enclosed magnetic flux of that segment~\cite{yeates2011generalized}. 
Scientists can select a value for minimum persistence as determined by the approximate noise level based on the simulation or the desired level of detail. This approach supports a physics-based, data-driven parameter selection for noise reduction.

Within the visualization community, the closest related work is that of Tricoche and Sanderson et al.~\cite{sanderson2010analysis, tricoche2011visualization}. Both the related work and our research address the identification of critical points and separatrices. However, Tricoche and Sanderson analyze the intersection of the magnetic field to a Poincar\'{e} plot. This examines the dual of the magnetic field and not, as with the research presented, the magnetic field itself. Additionally, Tricoche and Sanderson employ ridgeline formation, the Jacobian and other subjective approximation methods to identify features. This is a vastly different approach from the topology-based contour tree framework presented and does not use a parameter such as persistence to limit the impact of noise. 

The research presented is the result of a collaboration with physicists who are experts in reconnection; the described analysis tools are a result of an iterative software development process. Our contributions to the visual analysis of magnetic reconnection are:
\begin{itemize}
    \vspace{-0.5em}
    \item The application of a 2D topological algorithm, i.e. a contour-tree based segmentation, to accurately and objectively identify magnetic islands and reconnection points
    \vspace{-0.5em}
    \item The use of persistence as a physics-based parameter to reduce the effects of noise in visualizations
    \vspace{-0.5em}
    \item The identification of a new island generation process, resulting in an improved understanding of kinetic plasma simulations 
\end{itemize}

\section{Simulation}
\paragraph*{Setup:} We ran the simulation using VPIC \cite{Bowers_2008a, Bowers_2008b}. This PiC code solves the relativistic Vlasov-Maxwell equations and describes kinetic plasma physics. It is widely used by plasma physicists to study magnetic reconnection \cite{Bowers_2007, Yin_2008, Daughton_2009, Guo_2014, Guo_2015, Guo_2016, Guo_2019, Le_2019, Li_2019, Stanier_2019, Kilian_2020}.
We setup the simulation as follows: We perform the 2D simulation in the $x-z$~plane, with the electric current in the current sheet flowing into the $y$ direction~(first diagram of Fig.~\ref{fig:magRecDiagram}). This out-of-plane direction $y$ is assumed to be invariant, i.e. all derivatives $\partial / \partial y$ vanish identically.  The current sheet is initially force-free with magnetic field strength $B_0$ and initial thickness of five $d_\mathrm{e}$, where $d_\mathrm{e}$ is the electron inertial length that denotes the electron kinetic scale. 
The $x$ direction is periodic for fields and particles. The simulation domain is $250 d_\mathrm{e} \times 125 d_\mathrm{e}$~($L \times W$) and is well-resolved using $1024 \times 512$ cells, each with 400 electrons and 400 ions. 
To trigger reconnection, we add a long-wavelength perturbation 
in the center of the simulation domain. 
We save the magnetic field at regular intervals in time for further analysis. In post-processing, we use a spectral solver to convert the magnetic field $\vec{B}$ into the magnetic vector potential $\vec{A}$. More details on VPIC are included in the supplementary material.


\paragraph*{Periodic Boundaries:} Though magnetic and electric fields can theoretically extend to infinity, it is impossible to simulate on computers with finite resources. Physicists identify a rectangular, finite domain to simulate and achieve the effect of infinity by repeating the domain in a single or in both directions. This is reflected in the simulation's uni-periodic (left/right \textbf{\textit{or}} top/bottom periodic) or bi-periodic (left/right \textbf{\textit{and}} top/bottom periodic) setup. Therefore, the contour-tree results must account for this periodicity. This is accomplished by topologically deforming the uni-periodic data into 
an annulus and the bi-periodic data into a torus. This occurs during the triangulation phase, where the vertex-based, uniform quadrilateral grid is connected into a triangular mesh for analysis. The data is first triangulated by connecting diagonal vertices of each quadrilateral. Then, for each pair of boundaries to be stitched, there are multiple pairs of triangles added to join the first and last layers of vertices. 

Transforming a rectangle to 
an annulus or a torus converts the domain to a manifold where the saddle points are no longer fully defined by a contour tree. We determine using Euler's characteristic~\cite{Tierny:2017:Springer:Topological} that for both uni- and bi-periodic data, there are exactly two saddle points missing from the contour tree results, regardless of the complexity of the data itself. These two saddle points are subsequently identified using discrete Morse geometry~\cite{banchoff1970critical}, and in Figure~\ref{fig:teaser} right, shown as the red dots at the top-left and bottom-right corners of the figure. These saddle points correspond to a topological change in the genus rather than define component connectivity. However, the segmentation, minima and maxima determined by the contour tree are assured to be accurate and complete.
\section{Results}
\subsection{Objective Segmentation of Magnetic Islands}
We segment the triangulated scalar values of $A_\mathrm{y}$ using a standard contour tree pipeline in TTK. We first determine the minimum persistence threshold suitable for the noise given the simulation setup.  \textit{Topological Simplification} then simplifies the topology based on the restrictions of persistence. This minimizes the impact of PiC noise on the analysis results. The \textit{ftmTree} filter then computes the contour-tree based segmentation given the remaining topology. From these results, we extract the segments and critical points. We also use \textit{ScalarFieldCriticalPoints} to extract any remaining saddle points that exist as a result of a change in genus. Each vertex in the data is given a segment ID or designated as a critical point (with associated type: minimum,  saddle or maximum). This topological analysis is independently computed for every output time step.

\begin{figure}[b!]
\vspace{-2em}
  \centering
  \includegraphics[width=\columnwidth]{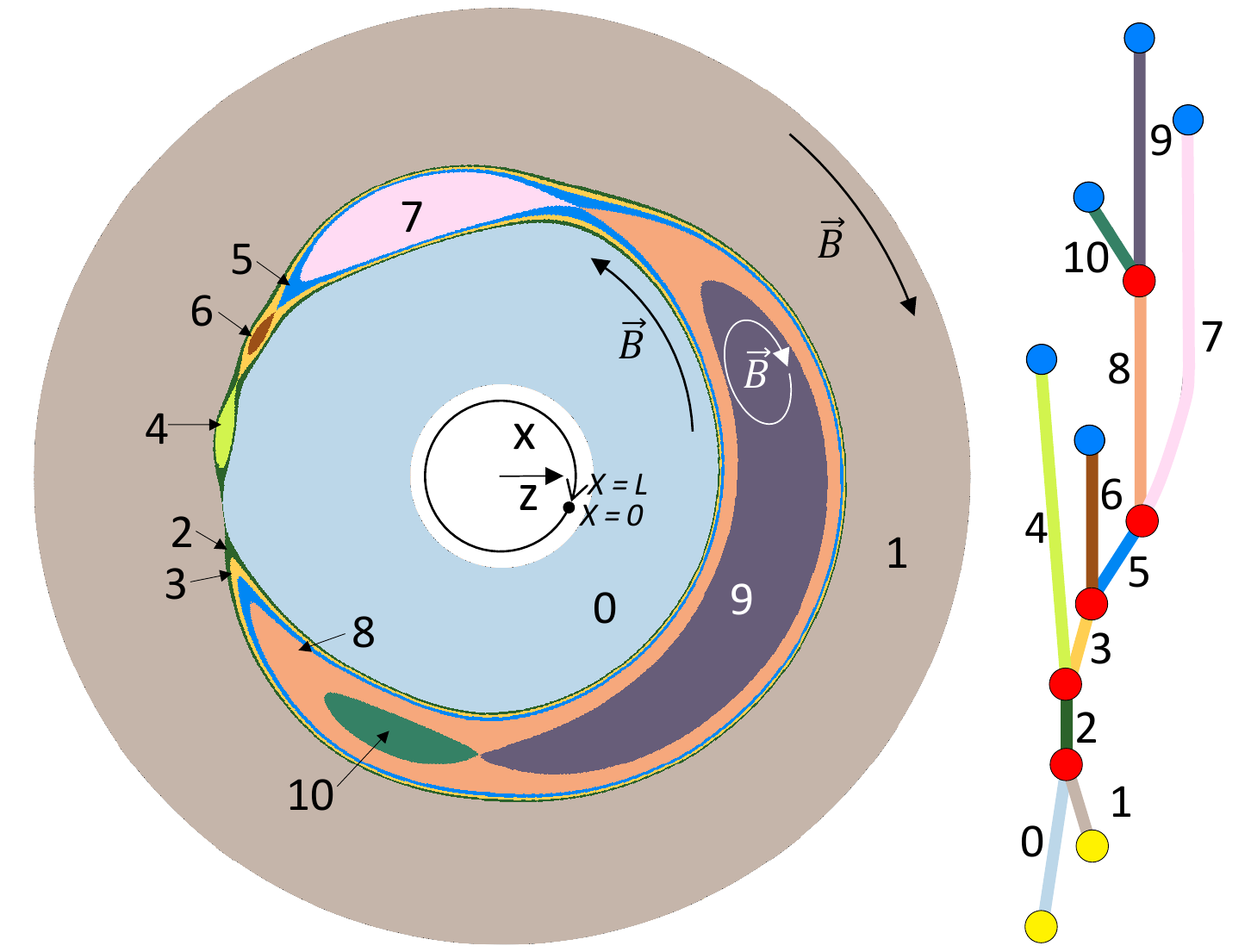}
  \vspace{-2em}
  \caption{\label{fig:ts510_periodicity}
     \textbf{The contour tree.} We show here the segmentation results for output step 510 at persistence $0.1 B_0 d_\mathrm{e}$ as an annulus and as a contour tree. Each color identifies a unique segment in the results. The arrows in segments $0$, $1$ and $9$ illustrate the direction of the magnetic field. The edges and vertices of the contour tree are in ascending order of $A_\mathrm{y}$ rather than proportional to $A_\mathrm{y}$, so that features with a small persistence are plainly visible. The nested hierarchy of the magnetic islands is also clearly visible, for example, segments $9$ \& $10$ are nested inside segment $8$.}
\end{figure}

The segmentation of output number 510 of the simulation at a persistence of $0.1 B_0 d_\mathrm{e}$ is shown as a rectangle in the right panel of Figure~\ref{fig:teaser} and as an annulus in Figure~\ref{fig:ts510_periodicity}. 
The persistence of $0.1 B_0 d_\mathrm{e}$ is a few times the amplitude of random fluctuations in the simulation, but only $0.0017$ of the total change in $A_\mathrm{y}$ at $t = 0$ and therefore removes only the smallest of segments.
In Figure~\ref{fig:ts510_periodicity}, both the annulus and contour tree identify segments uniquely by a color and a segment ID. As in the right panel of Figure~\ref{fig:teaser}, the vertices of the contour tree graph are colored yellow for minima, red for saddle points and blue for maxima. Note that in the contour tree, there are 6 maxima, 6 saddle points and 2 minima. Euler's characteristic gives us $6 - 6 + 2 = 2$; the characteristic for an annulus is $0$, meaning that we were missing 2 saddles. This confirms the limitations of the contour tree algorithm on periodic data.

Figure~\ref{fig:ts510_periodicity} also shows the direction of the magnetic field $\vec{B}$ in segments $0$, $1$ and $9$. The clockwise orientation illustrated in segment $9$ is representative of the orientation of all segments except for $0$ and $1$. Segments $0$ and $1$ are topologically unique because while the other segments are homeomorphic to a disc, segments $0$ and $1$ are homeomorphic to an annulus. Therefore, segments $0$ and $1$ are not considered to be magnetic islands in plasma physics. It is clear from the segmentation and contour tree that an empirical study of this data at one or a few isovalues of $A_\mathrm{y}$ would be insufficient to understand and characterize its complex hierarchy.



\begin{figure}[t!]
  \centering
  \includegraphics[width=\columnwidth]{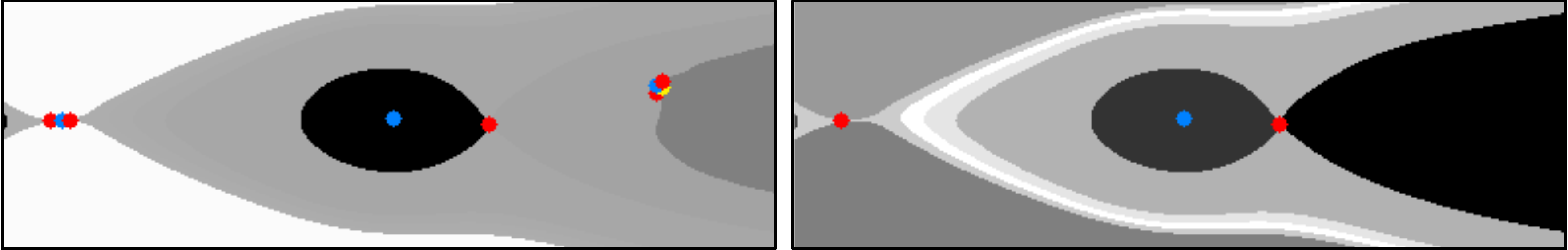}
  \vspace{-2em}
  \caption{\label{fig:ts510_persistence}
     \textbf{Effectiveness of persistence.} We show a magnified region of results for output 510 at persistence $0 B_0 d_\mathrm{e}$, left, versus $0.1 B_0 d_\mathrm{e}$, right. At persistence $0 B_0 d_\mathrm{e}$, in addition to noise at the $+z$ and $-z$ boundaries (not shown), noise within the segments exhibits as clusters of critical points. They are removed at persistence $0.1 B_0 d_\mathrm{e}$}
     \vspace{-1.5em}
\end{figure}

The importance of the persistence parameter is shown in Figure~\ref{fig:ts510_persistence}, with a magnified section of the results of output 510 at a persistence of $0 B_0 d_\mathrm{e}$ as compared to the results at the scientifically more meaningful value of $0.1 B_0 d_\mathrm{e}$. The noise in the system, shown in the left panel of Figure~\ref{fig:ts510_persistence} as clustered critical points at multiple locations, can lead to ambiguity when determining the location of reconnection points and generate superfluous segments in regions of interest. In total, the segmentation at persistence $0 B_0 d_\mathrm{e}$ generates an additional $858$ segments, $426$ minima, $429$ saddle points and $3$ maxima. 
  
\subsection{Detecting Lakes on Mountains}

\begin{figure}[b!]
   \vspace{-1em}
  \centering
  \includegraphics[width=\columnwidth]{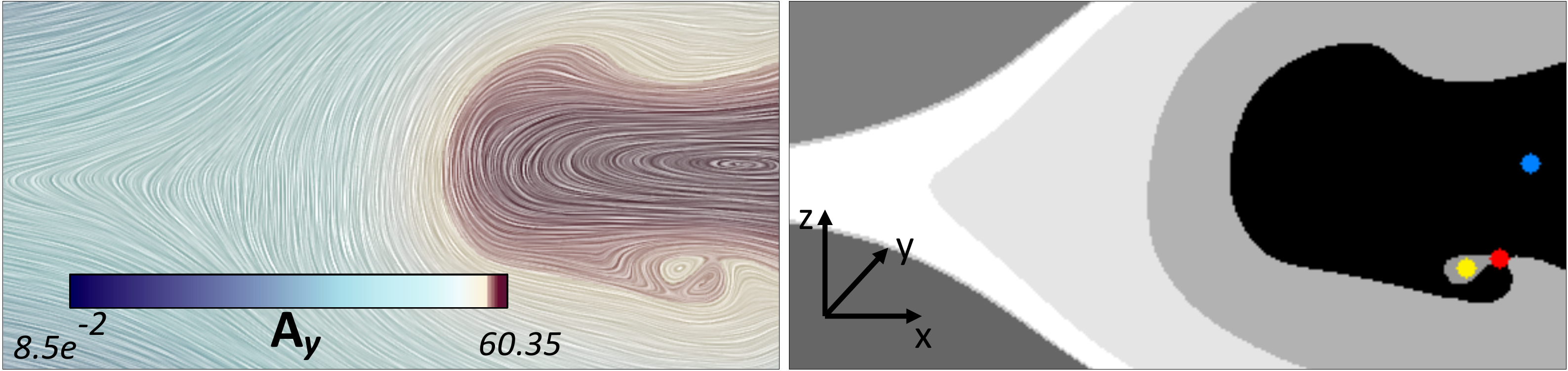}
    \vspace{-2em}
  \caption{ \label{fig:minAyResults}
    \textbf{Lake on the mountain.} A magnified region of output 693 highlights an island with a minimum enclosed by an island with a maximum, a unique feature not reported previously.}
\end{figure}
\begin{figure}[tb]
  \centering
  \includegraphics[width=\columnwidth]{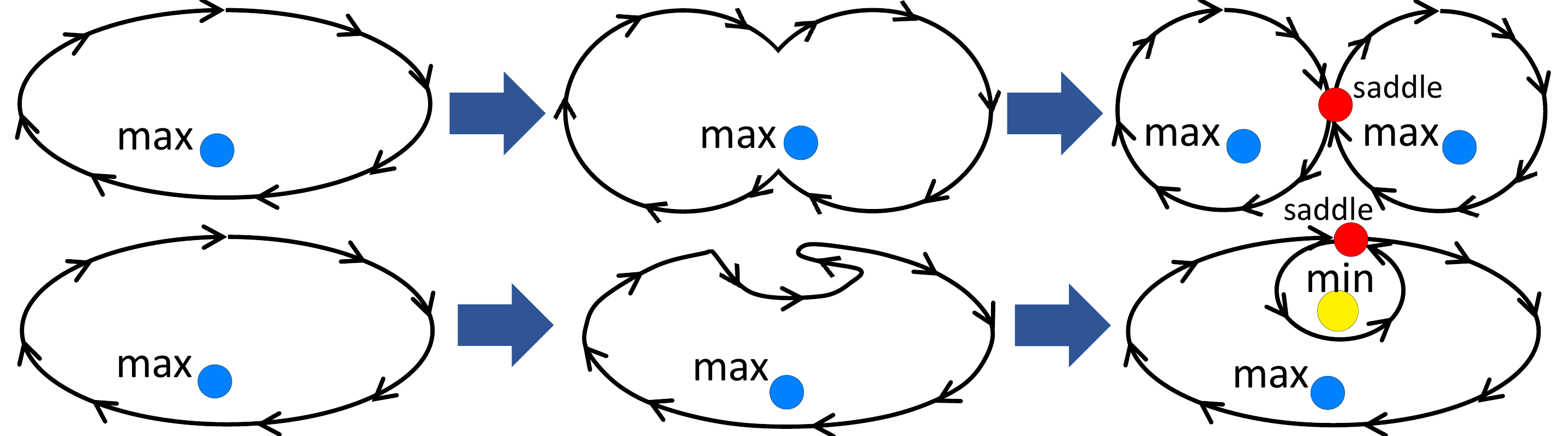}
    \vspace{-2em}
  \caption{\label{fig:minAyDiagram}
     \textbf{Two of a kind.} The top row illustrates the typical process by which new islands are formed during reconnection. The bottom row illustrates a new type of island formation discovered through the use of our contour-tree segmentation algorithm. Here, the charged particles press into the magnetic field, creating a depression. This depression is eventually pinched off, resulting in a region where the magnetic field points in the reverse direction. This region corresponds to a minimum critical point.}
     \vspace{-1.5em}
\end{figure}

Typically, when new islands form, the plasma kinetic effects dissipate the magnetic field inside the current sheet, causing magnetic field lines to bend toward one another and reconnect. 
In this process, the magnetic field line in question connects in the middle, going from an ``ellipse'' to a ``figure 8'' and preserving the original orientation of the magnetic field line. This is shown in the top row of Figure~\ref{fig:minAyDiagram}. The results of this process are seen in Figure~\ref{fig:ts510_periodicity} where the direction of the $\vec{B}$ field of segment $9$ aligns with segment $1$ on the outer border and aligns with segment $0$ on the inner border. This also applies to the other islands, meaning that they all have the same \textit{clockwise} magnetic field orientation and contain local maxima. However, through the use of the contour tree segmentation algorithm, we have identified a different process for island generation that has not been reported previously. 
The generation of these unique segments signifies a new type of magnetic field behavior. We observe that plasma particles push the magnetic field lines into a concave depression, resulting in a new island formation where one island is inside another. This is contrary to the more common behavior in reconnection where the magnetic field drives particle motion. This new process of island formation is diagrammed in the bottom row of Figure~\ref{fig:minAyDiagram}. These new inner islands correspond to a \textit{counter-clockwise} magnetic field orientation and contain local minima as critical points. A magnified region of output 693 of the simulation shown in the right panel of Figure~\ref{fig:minAyResults} identifies this feature as the segment containing a yellow dot. On the left, we use a custom colormap to enhance the visibility of the region of interest in the LIC visualization. The concentric circles and colors of the LIC at this region on the left coincide with the segment identified on the right. This confirms that the island detected is valid and not an artifact of our method. If data was scaled in the third dimension by $A_\mathrm{y}$, this new segment would be akin to a lake on the side of a mountain. 

This new process for island generation was first observed through the application of the contour-tree algorithm mainly due to the algorithm's accuracy and robustness in identifying features.
The segments with counter-clockwise magnetic field orientation tend to be very small compared to the overall domain, they occur infrequently in the midst of the simulation and appear only for a few time steps once formed. Therefore, it is highly likely that an empirical analysis of the data would have missed them. Such small segments might also have been misidentified as noise, if a parameter such as persistence wasn't used to restrict the influence of noise to the segmentation.



This unexpected discovery has motivated us to reexamine aspects of magnetic reconnection simulations to better understand what process led to these results. We have hypothesized that these events may signal a reversal in the magnetic energy conversion, i.e., from the plasma kinetic energy back into magnetic energy. However a deeper examination of the simulation must be conducted before these theories can be confirmed. 


The two processes for island generation described in this paper
are very different topologically and pose the question of what else we may discover through the continued application of topological analysis for magnetic reconnection. It is clear that understanding magnetic reconnection in 2D is ultimately a study in the way magnetic field lines bend, intersect and interact with each other. This is, after all, the essence of the study of topology itself.



\section{Evaluation}
\begin{figure}[h]
  \centering
  \includegraphics[width=.55\linewidth]{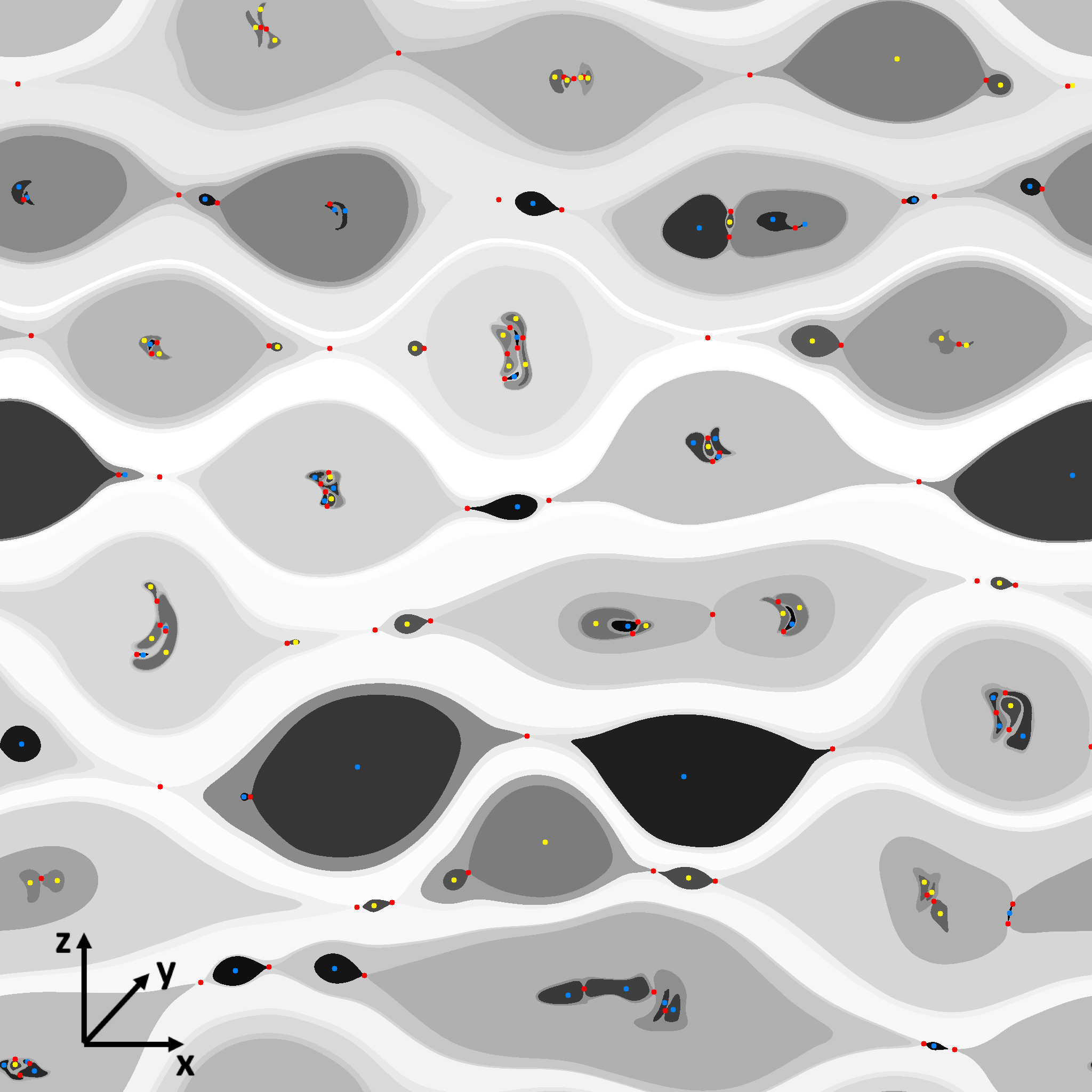}
  \includegraphics[width=.43\linewidth]{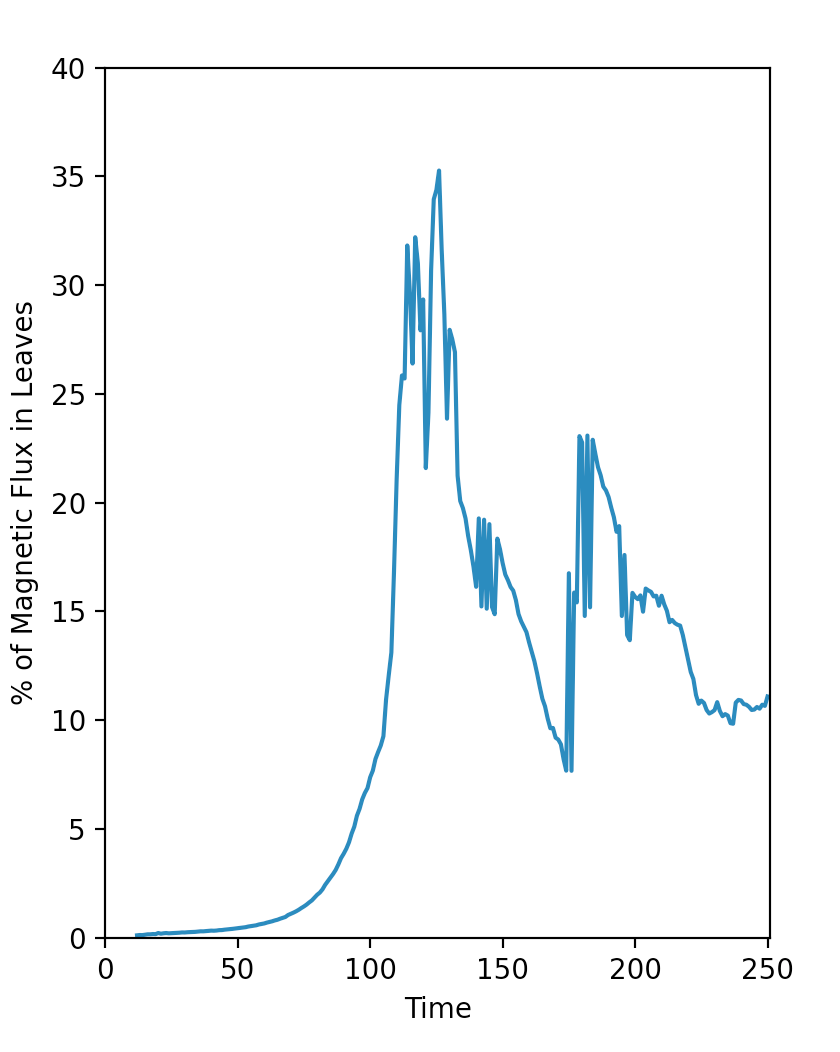}
    \vspace{-1em}
  \caption{\label{fig:8cs}
     \textbf{The real challenge.} We present a complex VPIC simulation with eight current sheets stacked vertically. At persistence $0.01 B_0d_\mathrm{e}$, this data is split into 179 segments, with 45 maxima, 46 minima, and 91 saddle points (including the two as a result of a change in genus).}
     \vspace{-1.5em}
\end{figure}

We have applied the analysis to more complicated simulations involving 8, 16 or 32 current sheets. These simulations contain many complex features of interest to physicists studying reconnection. A comprehensive evaluation of such data is impossible through traditional analysis techniques. In contrast, the presented contour-tree algorithm can efficiently and accurately identify relevant features.
Figure~\ref{fig:8cs} left shows an example of the application of this algorithm to a large PiC simulation with bi-periodic boundaries. The simulation domain resolution is $500 d_\mathrm{e} \times 500 d_\mathrm{e}$ in the $x-z$~plane, with $2048\times 2048$ cells~\cite{Du2020}. It starts with eight current sheets and becomes more turbulent over time. 
The complex nested hierarchy of the data, captured by the segmentation, makes it apparent why the described tool is valuable for understanding magnetic reconnection. In comparison, the Servidio et al. segmentation algorithm identifies only the leaves of the contour trees, i.e., the segments corresponding to maxima and minima. Figure~\ref{fig:8cs} right examines the combined magnetic flux in the leaf-segments of the contour tree as a percentage of the total. Of the 251 output steps, we disregard the first 12 output steps to allow time for reconnection events to start. The magnetic flux in the leaves over the following 239 outputs range from $0.10\%$ to $35.27\%$ of the total with a mean of $10.76\%$. It's clear that restricting the analysis of the data to just the leaves would lead to a limited understanding of reconnection events.

A comparison of the traditional isovalue-based approach, the approach by Servidio et al.~\cite{servidio2009magnetic, servidio2010statistics}, and our contour-tree based technique shows the advantages and disadvantages of each method. The traditional isovalue-based method can be useful for a quick initial exploration of the data, using an interactive tool such as Paraview~\cite{ayachit2015paraview}, commonly used by scientists.
However, for a more quantitative analysis, the contour-tree based algorithm is a better option. The contour-tree based algorithm is ``objective''; even the single threshold parameter is defined by simulation properties. In comparison, the traditional method is entirely subjective; the method described by Servidio et al. is subjective when categorizing critical points as physically valid or noise. 
The contour-tree based algorithm also produces the entire hierarchy of the data and the corresponding contour tree, making it possible to recognize correlations between segments (including nested islands), separatrices and critical points.
This allows for more informed physics-based conclusions as compared to other methods. 
Finally, every element captured in the contour-tree based segmentation has an corresponding definition in kinetic plasma physics; the two are fundamentally consistent with one another.

Our physicist co-authors have stated that ``Cutting-edge research in magnetic reconnection must understand the roles of nonlinear structures in energy conversion, heating, and particle acceleration. Recent advances in computational plasma simulations on peta-scale supercomputers have enabled us to model hundreds or even thousands of segments in one simulation. However, to quantify the effects of magnetic islands and X points, one needs an efficient and robust way to identify those structures. The contour-tree based framework is extremely helpful for accomplishing these goals. Additionally, because the entire tool is built on an open-source framework, collaboration between colleagues is more productive as one can focus on the physics rather than the particulars of the analysis method. Finally, the tool provides information about each segment, separatrix or critical point individually. This relevant output can be used to compute additional statistics such as size distributions, particle energy spectra and acceleration mechanism related quantities.''

\section{Conclusion}
We show the successful application of the contour-tree based technique for reconnection study given the one-to-one correspondence between 2D magnetic reconnection and scalar field topology.
In addition, our use of persistence to reduce the impact of noise is highly advantageous when compared to other analysis techniques. 
Our work has offered physicists a much more robust way to examine 2D magnetic reconnection and provided a new perspective of the underlying mechanisms. 
In future work, we plan to add a suite of statistics tools to automatically extract physically-relevant information. Next steps also include temporal tracking of islands and reconnection points.

\acknowledgments{
We thank David Rogers, Jim Ahrens, Francesca Samsel and Hui Li. Funded by DOE/OFES, LANL LDRD and LANL Cinema Project. 
}

\newpage
\FloatBarrier
\clearpage
\newpage

\bibliographystyle{abbrv-doi}

\bibliography{template}
\end{document}